# Imaging the Long Transport Lengths of Photo-generated Carriers in Oriented Perovskite Films

Shuhao Liu,[†] Lili Wang,[‡] Wei-Chun Lin,[‡] Sukrit Sucharitakul,[†] Clemens Burda,*[,‡] and Xuan. P. A. Gao*[,†]

[†]Department of Physics, Case Western Reserve University, 10900 Euclid Ave., Cleveland, Ohio, 44106, United States
[‡]Department of Chemistry, Case Western Reserve University, 10900 Euclid Ave., Cleveland, Ohio, 44106, United States

*Supporting Information Placeholder*

**ABSTRACT:** Organometal halide perovskite has emerged as a promising material for solar cells and optoelectronics. Although the long diffusion length of photo-generated carriers is believed to be a critical factor responsible for the material's high efficiency in solar cells, a direct study of carrier transport over long distances in organometal halide perovskites is still lacking. We fabricated highly oriented crystalline $CH_3NH_3PbI_3$ ($MAPbI_3$) thin film lateral transport devices with long channel length (~ 120 μm). By performing spatially scanned photocurrent imaging measurements with local illumination, we directly show that the perovskite films prepared here have very long transport lengths for photo-generated carriers, with a minority carrier (electron) diffusion length on the order of 10 μm. Our approach of applying scanning photocurrent microscopy to organometal halide perovskites may be further used to elucidate the carrier transport processes and vastly different carrier diffusion lengths (~ 100 nm to 100 μm) in different types of organometal halide perovskites.

The rapid improvement of perovskite solar cells' efficiency has been remarkable: certified solar cell efficiency has reached 22.1% within just a few years after the initial work.[1-9] The current high efficiency organometal halide perovskite (OMHP) solar cells take advantage of ambipolar carrier-transport ability of OMHP and have planar structure in which the photo-generated carriers diffuse over a short (< 1 μm) distance along the vertical direction and get collected by the hole transport material and the electron transport material.[3-6] Devices similar to the planar solar cell were studied to extract the carrier diffusion length which was found to range from 100 nm to ~ 1 μm.[10-15] It is also reported that large single crystals of triiodide perovskite have improved carrier mobility and lifetime, giving rise to estimated carrier diffusion length as long as >175 μm.[16] More recently, spatially scanned photoluminescence imaging microscopy directly showed carrier diffusion over ~ 10 μm distance in single-crystal OMHP nanostructures.[17] Theses outstanding properties of perovskite are as well found useful in optoelectronic applications.[18-23] In this work, we apply the scanning photocurrent microscopy (SPCM) to electrical transport devices of highly oriented $MAPbI_3$ film. The long lateral channel length (ca. 120 μm) and Schottky barriers at the contacts allowed us to selectively quantify the minority carrier (electron) transport length and the spatially resolved imaging of photocurrent gave a direct map to visualize the carrier transport characteristic variations over large sample area.

The studied $MAPbI_3$ films were spin-coated[24] and thoroughly characterized for high phase purity and crystallinity. One promising recent addition that has been carefully studied is the addition of chloride, in the form of $PbCl_2$ or $CH_3NH_3Cl$ (MACl), to the perovskite precursor solution.[13,25] The resulting perovskites have shown improved solar cell performances. Figure 1a shows the crystal structure of OMHP. In our experiment, two electrodes made of Au or Ni were first deposited onto the insulating silicon oxide surface of a silicon wafer. The two metal electrodes had a size of 8 mm × 8 mm separated by a gap (channel) of ~ 120 μm width. The $MAPbI_3$ film (~ 300 nm thick) was subsequently deposited onto the whole wafer covering both electrodes and the channel. Immediately after the perovskite film deposition, a layer of parylene (~ 1 μm thick) was coated onto the chip to encapsulate perovskite and prevent sample degradation. As Figure 1b shows, X-ray diffraction (XRD) of these films proves high crystallinity and orientation along the (110) plane. Scanning electron microcopy (SEM) images reveal a polycrystalline character of the film, with lateral grain sizes of ~ 0.5-1 μm as shown in Figure 1c.

In our SPCM measurements, light from a halogen white light source or a He-Ne laser (wavelength $\lambda$ = 633 nm) was focused onto the device via the objective of a microscope (numerical aperture NA = 0.8) to generate electrons and holes moving towards the source (S) and drain (D) electrodes (Figure 2a). Photocurrent measured through the circuit originates from those photo-generated carriers that reach the electrodes and go through the perovskite-metal junction. If Schottky barrier exists at the contacts, the minority carrier's transport length is the key factor limiting the photocurrent and can be obtained via analyzing the photocurrent as a function of the separation between the light spot and channel edge.[26-29] Furthermore, the photocurrent is recorded while the device is scanned along X- and Y-directions, yielding a two-dimensional (2D) map of carrier diffusion/transport characteristics.

We first characterize the current-voltage (IV) characteristics and establish the metal-perovskite junction as Schottky contact. Under the global illumination condition (Figure 2b inset), both Au and Ni yield Ohmic contacts (Figure 2b shows an example IV for a device (device #1) with Au contacts). However, when only one of the metal-perovskite contact junctions is illuminated, the IV curves become asymmetric, as illustrated in Figure 2c and d for device #2 with Au contacts, and device #3 with Ni contacts. We also note that such asymmetry (or rectifying behavior) in IV is generally more prominent in devices with Ni contacts. This behavior resembles other metal/semiconductor-nanostructure/metal devices[26-28] and the asymmetric IV here is understood by viewing the two contacts as two back to back connected diodes (Figure 2d inset). The perovskite/metal junction behaving like a Schottky diode with the perovskite as the anode suggests that our perovskite is p-type, in agreement with literature about holes being the majority carriers in such OMHP.[13,16] Moreover, the fact that devices with Au contacts show weaker asymmetry in the IV curves under local illumination further corroborates the p-type nature of perovskite given the larger work function of Au. The p-type nature of $MAPbI_3$ films here was also directly confirmed by the sign of thermoelectric-voltage (Figure S1, Supporting Information).

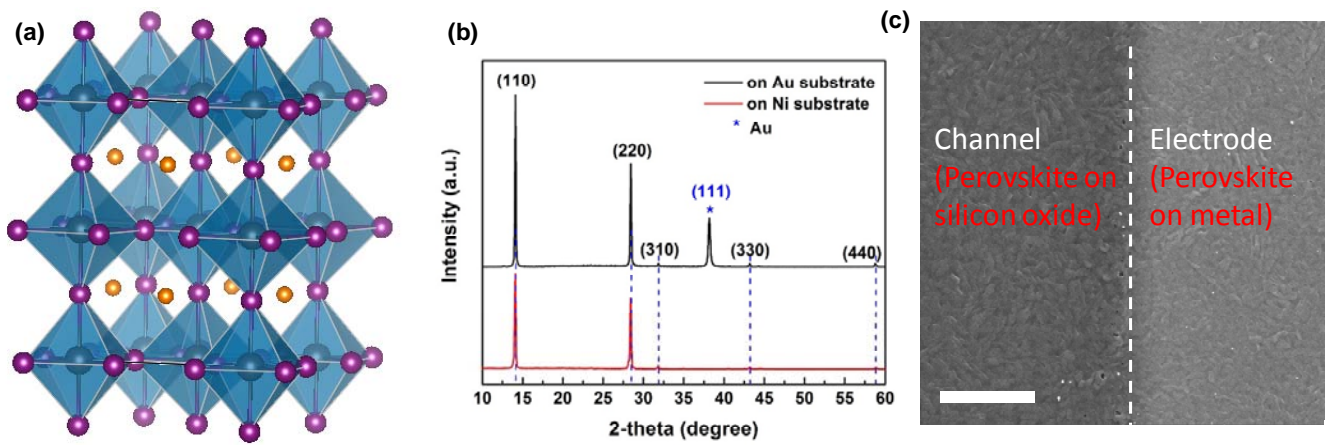

**Figure 1.** MAPbI$_3$ Perovskite. (a) Structure of organometal halide perovskites. Yellow sphere: CH$_3$NH$_3^+$ group; purple sphere: halide atom. The metal atom resides in octahedron center. (b) XRD patterns of prepared tetragonal phase MAPbI$_3$ perovskite films on substrates with Au (black curve) or Ni electrodes (red curve), respectively. The (111) peak originates from Au electrodes. (c) SEM image of a lateral perovskite transport device near one of the electrodes. The white dashed line marks the edge of electrode. The scale bar is 5 μm long.

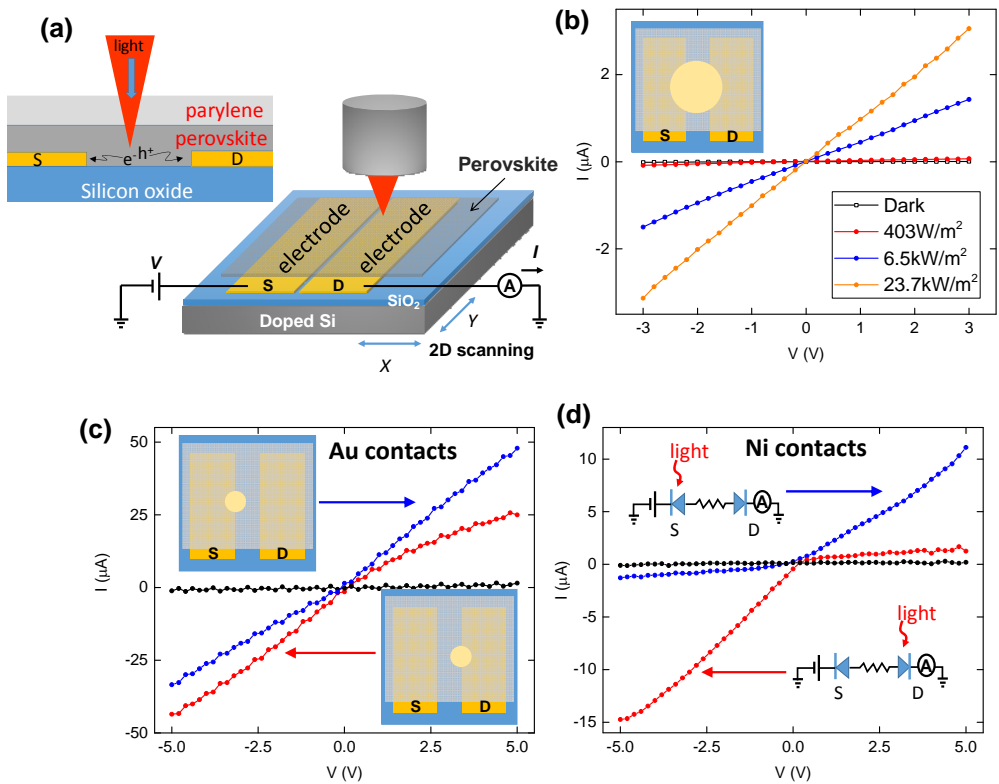

**Figure 2.** Experimental setup and contact IV characterization. (a) Schematic setup of scanning photocurrent experiment. (b) IV curves for device #1 with Au electrodes, where both contacts were illuminated at different white light intensity. The gap between electrodes was 124 μm wide and illumination spot diameter was 550 μm, covering both contacts and channel, as shown in the inset. (c) IV curve for device #2 with Au electrodes. (d) IV curve for device #3 with Ni electrodes. In (c) and (d), devices were under local illumination during measurement, either source contact (blue lines) or drain contact (red lines) was illuminated, spot diameter: 180 μm for (c) and 120 μm for (d), half of the illumination spot is inside the gap. Illumination intensity: 3.2 kW/m$^2$, black lines show dark current. Insets in (c) illustrate illumination configuration. Insets in (d) show the equivalent circuit of the devices.

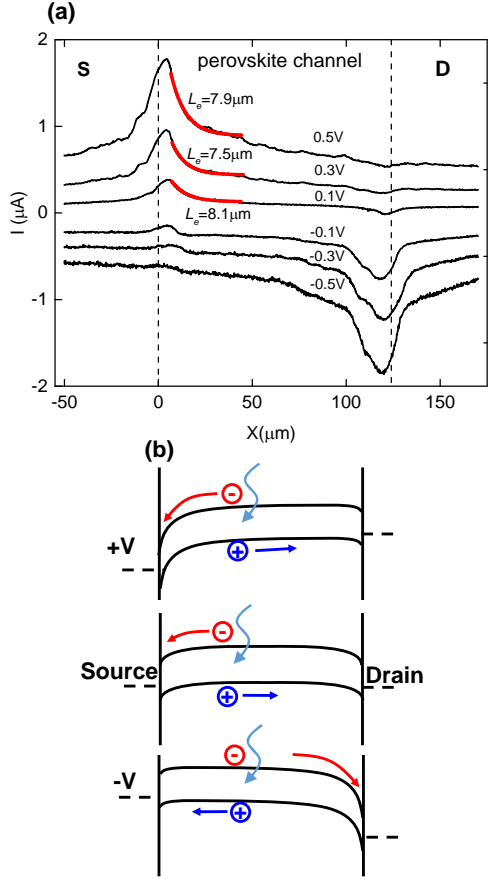

**Figure 3.** Photocurrent line-scans and device band diagram. (a) Photocurrent line-scans on device #1 with a 124 μm wide channel under various biases using focused laser illumination. Applied bias on the source is shown on top of each curve. Scans have been offset for clarity. Two dashed lines indicate the electrode edge for the source electrode and drain electrode separately. X is defined as the distance between laser spot and source electrode edge. (b) Device band diagrams. From top to bottom, applied bias on source is positive, zero and negative. In (a), relaxation length of electron current is extracted by fitting curves with exponential function.

We now turn to the SPCM results to discuss the carrier transport mechanism and diffusion length. To obtain a spatial decay profile of diffusing carriers and address the influence of drift current in the presence of strong electric field, we performed SPCM using focused laser light with illumination spot size ~ 2 μm (intensity ~1 kW/cm$^2$) as shown in Figure 3a for device #1 under bias from -0.5 V to +0.5 V (photocurrent line scan profiles taken with light excitation with larger spot sizes are shown in Figure S2, Supporting Information). When the laser spot is scanned across the device with +0.5 V applied between the source and drain, a dominant photocurrent peak appears at the edge of the source electrode. At smaller bias voltages, e.g. +0.1 V, the positive photocurrent peak at the edge of source electrode becomes weaker but a negative dip starts to appear near the edge of drain electrode. As the magnitude of the negative bias increases, this negative dip grows and the peak at the source electrode edge becomes more suppressed. This set of behavior has also been found in the SPCM of other semiconductor nanodevices[27] and can be understood by the combined effect of external and built-in electric fields at the Schottky contacts[27,29]. We use the schematic band diagrams of device under different bias conditions in Figure 3b to explain the data in Figure 3a (see Figure S3a in Supporting Information for additional data). Under zero bias voltage (Figure 3b, middle panel), the band diagram is symmetric about the center of device, but the built-in electric fields at the source and drain contacts point to opposite directions, causing a positive (negative) photocurrent when the photo-excitation is located near the edge of source (drain) electrode. As the illumination spot moves away from contact, the number of electrons that can diffuse to the source contact drops exponentially according to the electron diffusion length $L_e$ and gives rise to an exponentially suppressed photocurrent. Thus a peak (dip) is expected to be observed at the edge of the source (drain) contact. When a positive bias voltage is applied (Figure 3b, top panel), the voltage drop is mainly taken by the reversely biased source Schottky contact which now has stronger band bending and more efficient electron-hole separation, giving rise to larger photocurrent (a stronger peak) near the source electrode. Note that in Figure 3b, the width of Schottky barrier (i.e. the depletion width) where the build-in electric field exists is exaggerated for clarity.

With bias voltage between ±0.5 V, fitting the line-scan profile on main peak to exponential decay (red lines in Figure 3a) yields decay lengths ~ 7-8 μm, independent of the bias voltage.[30] This decay length corresponds to the electron diffusion length $L_e$. Additional study of photocurrent decay length at higher bias revealed an increasing trend of the decay length, reflecting the increased drift effect when there is a significant electric field inside the perovskite channel (Figure S3b and S4b, Supporting Information). Generally, we found that drift effects are small and the exponential fitting of spatial decay of photocurrent produces the diffusion length $L_e$ when the source-drain bias was limited to less than 0.3-0.5 V in our experiments.

2D SPCM images are obtained to elucidate the microscopic variation of carrier transport in perovskite films over large area. An image of ~ 200 μm × 150 μm area is shown in Figure 4a for device #4 with 0 V bias applied between source and drain. In Figure 4a, the color represents the magnitude of photocurrent and the edges of perovskite channel are marked as two dashed lines. First, there are two dominant photocurrent peaks with opposite sign at the edges of perovskite channel (shown as red and blue stripes). This double-peak feature in the SPCM image near the contacts reveals the Schottky nature of the contacts over all the area surveyed, although the magnitude of photocurrent peak varies slightly at different locations, likely due to local details of Schottky contact. Second, since the spatial decay of photocurrent near the contact edge is determined by $L_e$, the color gradient gives direct visualization of electron diffusion length. We see that in this SPCM image, appreciable color change happens over length scale on the order of 10 μm near the channel edges, indicating the characteristic $L_e$ being about 10 μm. In Figure 4b we summarize the statistical distribution of electron diffusion length fitted from line-scans in Figure 4a, most of counts fall within 8-14 μm for this device (see Figure S4 and Table S1 in Supporting Information for data on additional devices).

An electron diffusion length ≥ 10 μm is quite remarkable for perovskite film given that the grain size in the film is about 0.5-1 μm. Figure 4b gives an average $L_e$ = 10.5 ± 1.6 μm. These values obtained via SPCM in long channel lateral devices here are larger but not unreasonable when compared with values of estimated carrier diffusion length (~ 1 μm or larger) in $CH_3NH_3PbI_{3-x}Cl_x$ films in prior experiments.[10,14,15] We suggest that employing the electrical transport device with long channel length may be the reason that carrier transport length significantly longer than 1 μm can now be clearly seen here. In addition, the highly crystalline sample with predominant (110) orientation may also play a role in rendering long carrier transport length. Future studies on how the diffusion length variations correlate with the structural quality, material composition and stoichiometry variations will be very useful to elucidate the origin of the observed diffusion length variation over different locations of the sample.

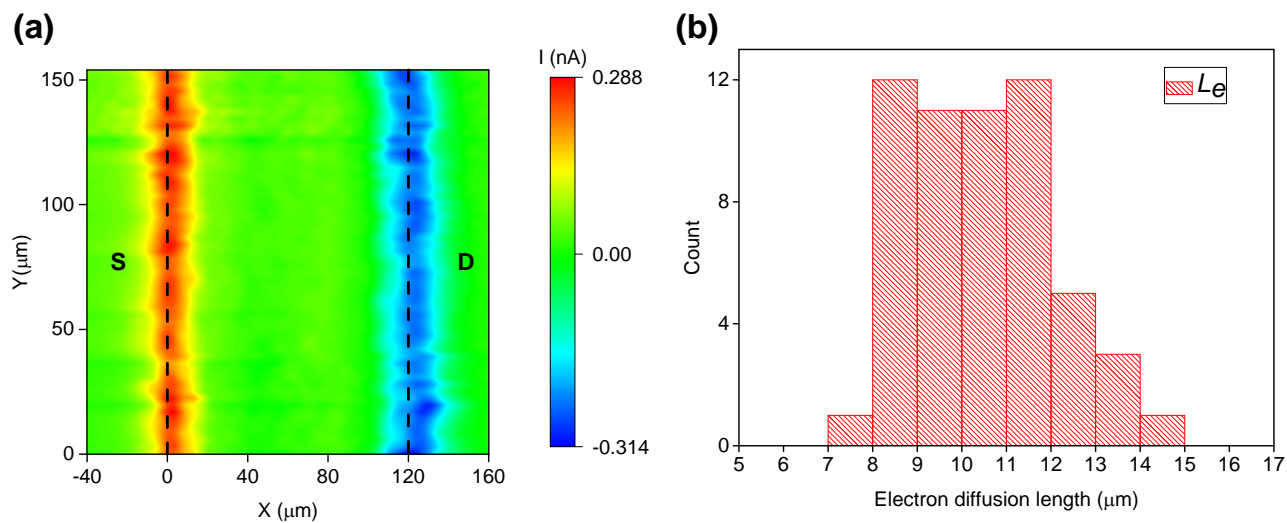

**Figure 4.** 2D photocurrent mapping and electron diffusion length variation. (a) 2D scanning photocurrent image of device #4 under 0 V bias on source electrode. Two dashed lines indicate electrode edges. (b) Histogram of electron diffusion length $L_e$ extracted from different line-scans in (a), yielding average $L_e$ = 10.5 ± 1.6 μm.

In conclusion, scanning photocurrent experiments directly revealed a long carrier transport length in long channel electrical devices of perovskite thin films, without the need of knowing carrier mobility or lifetime. In addition, our work offers further understanding of the interface between the perovskite and metal contacts in a simple electrical transport device. Via 2D SPCM, we were able to show spatial variations of charge transport characteristics over a large film area. These findings provide insights into both the fundamental carrier transport processes in halide perovskites and future device structure optimization.

## ASSOCIATED CONTENT

### Supporting Information

Materials and methods, thermo-electric voltage measurement, larger spot size illumination scanning photocurrent measurement, relation between applied bias and fitted decay length, 2D photocurrent mapping on another perovskite device, laser spot diameter measurement, SPCM on $MoS_2$ nanoflake, comparison between a single exponential function and a convolution integral of Gaussian function with a single exponential function, and a list of studied devices (including Figure S1-S7 and Table S1). The Supporting Information is available free of charge on the ACS Publications website.

## AUTHOR INFORMATION

**Corresponding Author**
*E-mail: burda@case.edu (C.B.); xuan.gao@case.edu (X.P.A.G.)
**Notes**
The authors declare no competing financial interests.

## ACKNOWLEDGMENT

X.P.A.G. thanks the National Science Foundation for its financial support under Award DMR-1151534, and Walter Lambrecht and Yanfa Yan for discussions.

*Supporting Information for*

**Imaging the Long Transport Lengths of Photo-generated Carriers in Oriented Perovskite Films**

Shuhao Liu,[†] Lili Wang,[‡] Wei-Chun Lin,[‡] Sukrit Sucharitakul,[†] Clemens Burda,*[,‡] and Xuan. P. A. Gao*[,†]

[†]Department of Physics, Case Western Reserve University, 10900 Euclid Ave., Cleveland, Ohio, 44106, United States
[‡]Department of Chemistry, Case Western Reserve University, 10900 Euclid Ave., Cleveland, Ohio, 44106, United States

**Materials and Methods**

*Substrate preparation*: Commercial silicon wafers (University Wafer) with 300 nm thick thermally oxidized $SiO_2$ layer were cut into around 1.5 cm × 1.5 cm pieces as substrates. These substrates were cleaned by sonication in acetone for 20 minutes, and rinsed sequentially with ethanol as well as de-ionized water, then blow dried. Two Ni electrodes or Cr/Au electrodes were then deposited onto substrates by e-beam evaporation or thermal evaporation (metal thickness ~ 50 nm). The electrodes had rectangle shape (~ 8 mm×6 mm) with a spacing around 120 μm which was created by using silver wire with 120 μm diameter as the shadow mask during metal deposition process. Before perovskite film deposition, substrates with electrodes were cleaned again with acetone, ethanol and de-ionized water, then blow dried.

*Perovskite film deposition and encapsulation*: The $MAPbI_3$ perovskite was deposited onto as-prepared substrates via spin-coating as described in previous literature.[1] The perovskite precursor solution was prepared with 1:1:1 molar ratio of $CH_3NH_3I$, $PbI_2$ and $CH_3NH_3Cl$ in dimethylformamide (DMF). After spin coating, the film samples were heated on a hot plate to promote crystal growth of perovskite. The process was performed in a nitrogen-filled glovebox. The resulted $MAPbI_3$ perovskite film thickness was ~ 300 nm as measured by AFM. Right after the perovskite film deposition, samples were transferred to a tube furnace and deposited with parylene-C to protect the perovskite film from being exposed to ambient atmosphere environment which may cause degradation effects. During the parylene deposition process, samples were kept at room temperature under 13.3 Pa pressure.

*Structure characterization*: Crystal structure of as-prepared samples were characterized and analyzed by using a Rigaku MiniFlex X-ray powder diffractometer using Cu Kα radiation ($\gamma$ = 0.154 nm) and a TESCAN SEM was used to image the sample.

*Current-voltage (IV) measurement and scanning photocurrent measurement*: For electrical measurement, part of the films of the sample was wiped away with ethanol soaked cotton swab to expose the metal electrodes for wiring. A National Instrument PCI-6221 DAQ card integrated computer along with a BNC-2090 terminal block and a Stanford Research Systems SR-570 low noise current preamplifier were used to supply bias and measure current. A LabVIEW program was used to record data. Sample was fixed on two stacked Newport AG-UC2 scanning stages and placed under the objective lens of a Zeiss Imager.A1m microscope for 2D XY scanning and mapping of photocurrent. A HAL 100 lamp integrated in the microscope with adjustable intensity or a Uniphase He-Ne laser coupled to the microscope was used as the optical excitation source. A shutter in the microscope was used to adjust the spot size when the Halogen light source was used. With careful alignment and a 50× objective lens (NA = 0.8) used, the laser spot was focused to be under 2 μm in diameter (Figure S5). 2D SPCM on different devices yielded a qualitatively similar photocurrent map (see Figure S4c for a 2D photocurrent map from another device in addition to Figure 4a in the main text). The SPCM was applied to a multi-layer molybdenum disulfide ($MoS_2$) device which showed ~ 2 μm decay length in the photocurrent line-scan (Figure S6). This control experiment further corroborates that the long decay length found in SPCM of $MAPbI_3$ films is not limited by the resolution of our SPCM setup. Table S1 summarizes the device contact types, measurements performed and electron transport lengths extracted from fitting the photocurrent decay for all the devices studied.

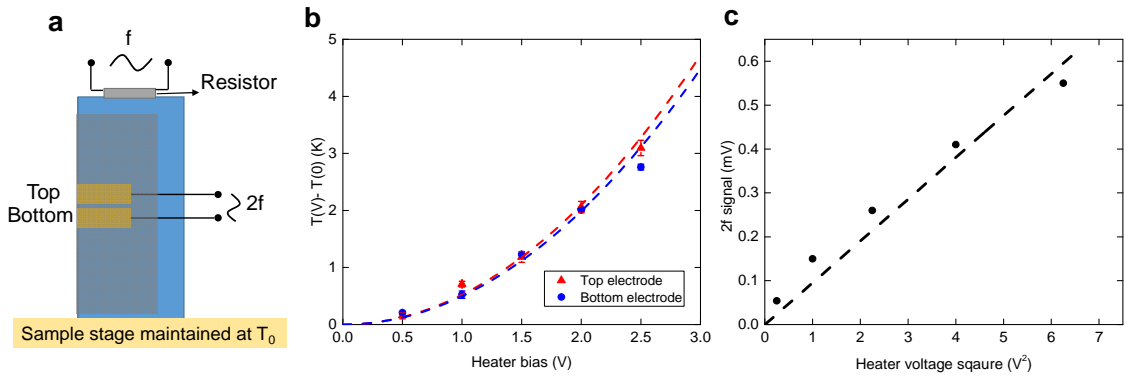

**Figure S1.** Thermoelectric-voltage measurement of MAPbI$_3$ film. (a) Schematic of the setup. A heater resistor was glued on top of the Si/SiO$_2$ substrate coated with MAPbI$_3$ film, for generating temperature gradient within the perovskite film. Sinusoidal AC bias with frequency $f$ was applied on the heater resistor to generate temperature gradient at frequency $2f$ with 90-degree phase shift due to Joule heating. Thermoelectric-voltage signal at $2f$ frequency and with 90-degree phase was detected between top electrode and bottom electrode (100 nm thick Ni, 120 µm apart) underneath the perovskite film. (b) Temperature difference between electrodes and thermal bath versus bias applied to heater resistor. The dashed lines show the $\Delta T \sim V^2$ dependence. (c) Thermoelectric-voltage signal at $2f$ frequency versus heater power. From these thermoelectric-voltage measurements, the perovskite samples can be determined to be p-type semiconductors.

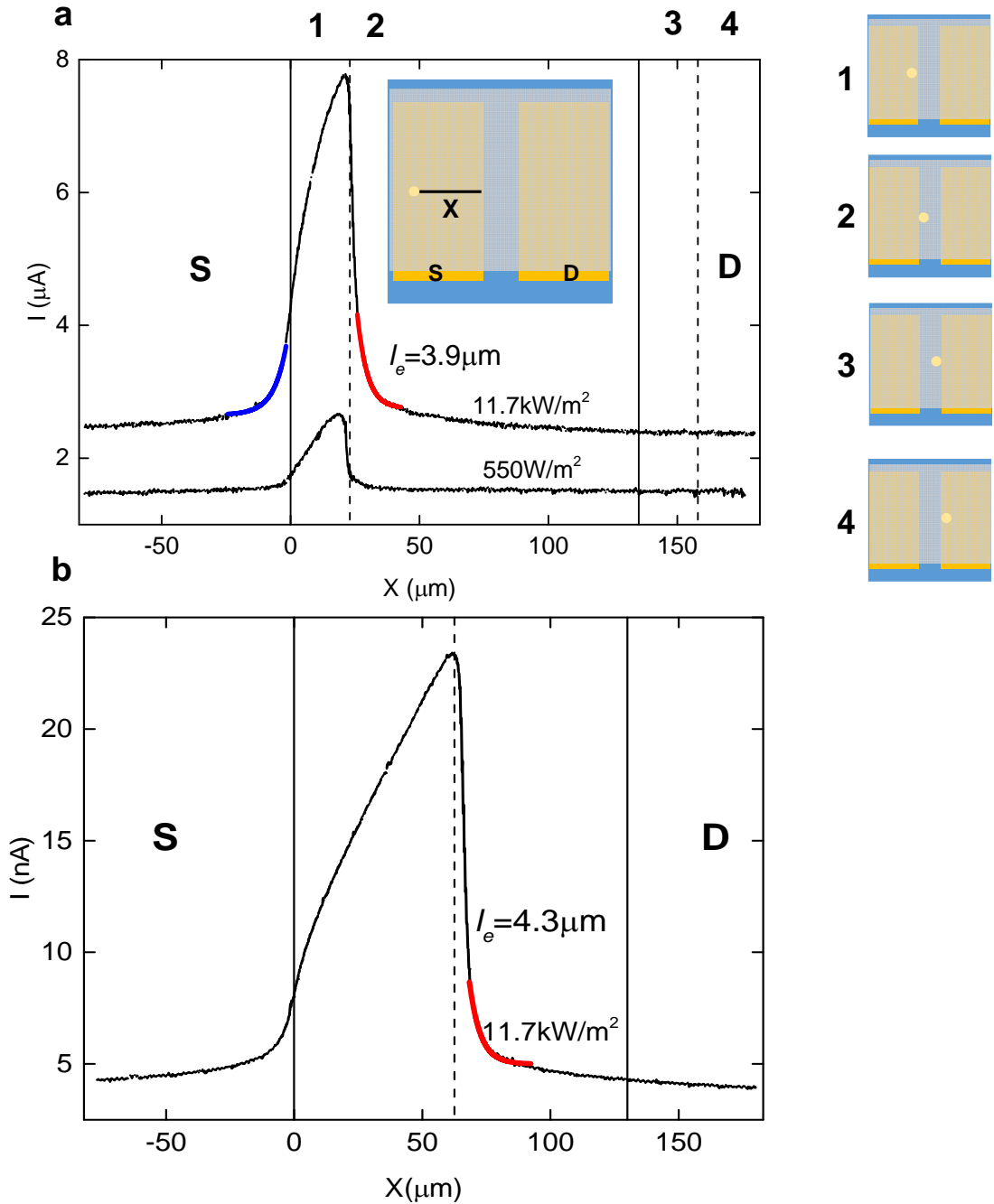

**Figure S2.** Photocurrent line-scans with finite size illumination spot. (a) Photocurrent line-scans with a 25 μm diameter white light illumination spot on device #5 with Au contacts in which the perovskite channel length was 135 μm and bias on source electrode was 5V. Inset shows the definition of X, which is the distance between spot front and source electrode edge. Diagrams to the right show the four important positions as the illumination spot scans across device which are also marked on the photocurrent versus position plot in the main panel. At X < 0, the focused light spot was completely outside the channel and on the source electrode where the electric field was nearly zero. The measured photocurrent relied on electrons/holes diffusing into the channel area. This diffusion limited photocurrent gave rise to an exponential (blue line) increase of current before the light spot's front started entering the channel (position 1 at X = 0). As the spot was moved further into the channel, the photocurrent showed a slower than exponential increase with the position until the whole spot has entered the channel such that there was no overlap between light spot and source electrode (position 2). The increase of current between position 1 and 2 reflects the fact that current measured under such a light/contact overlapping situation is set not by the diffusion ability of carriers but rather the density of photo-generated carriers. As the light spot moves beyond position 2 further into the channel, the current dropped rapidly, following an exponential decay (red line). (b) Photocurrent

line-scan with a 70 μm light spot as excitation source on the device #5, bias on the source electrode is 5 V. The light spot overlaps with the edge of the contact between X = 0 and 70 μm. Illumination intensities and fitted decay lengths for electrons are marked by line-scans in (a) and (b).

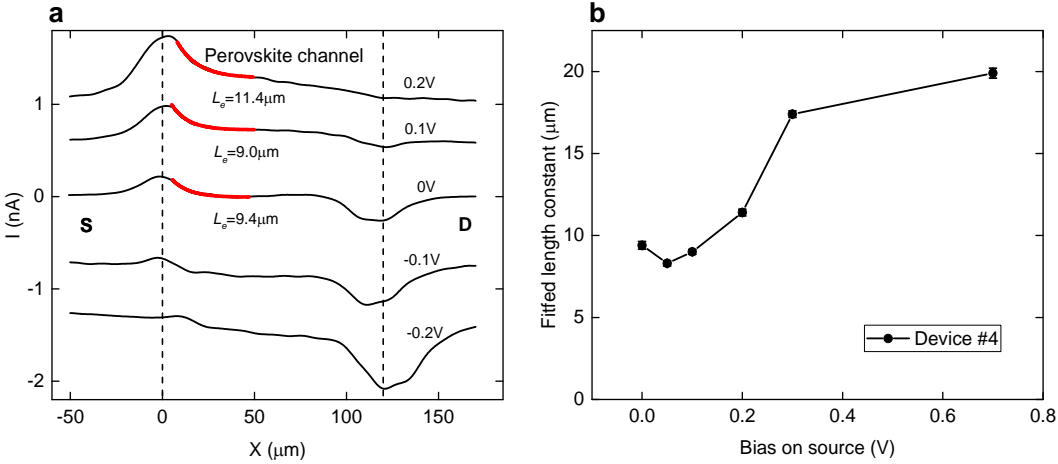

**Figure S3.** Photocurrent line-scans and bias dependent photocurrent decay length results for device #4 with 80nm Ni electrodes. (a) Photocurrent line-scans under various biases using focused laser illumination. Applied bias on source is shown on top of each curve. Two dashed lines indicate the electrode edge for the source electrode and drain electrode, separately. (b) Fitted decay length constant versus applied bias.

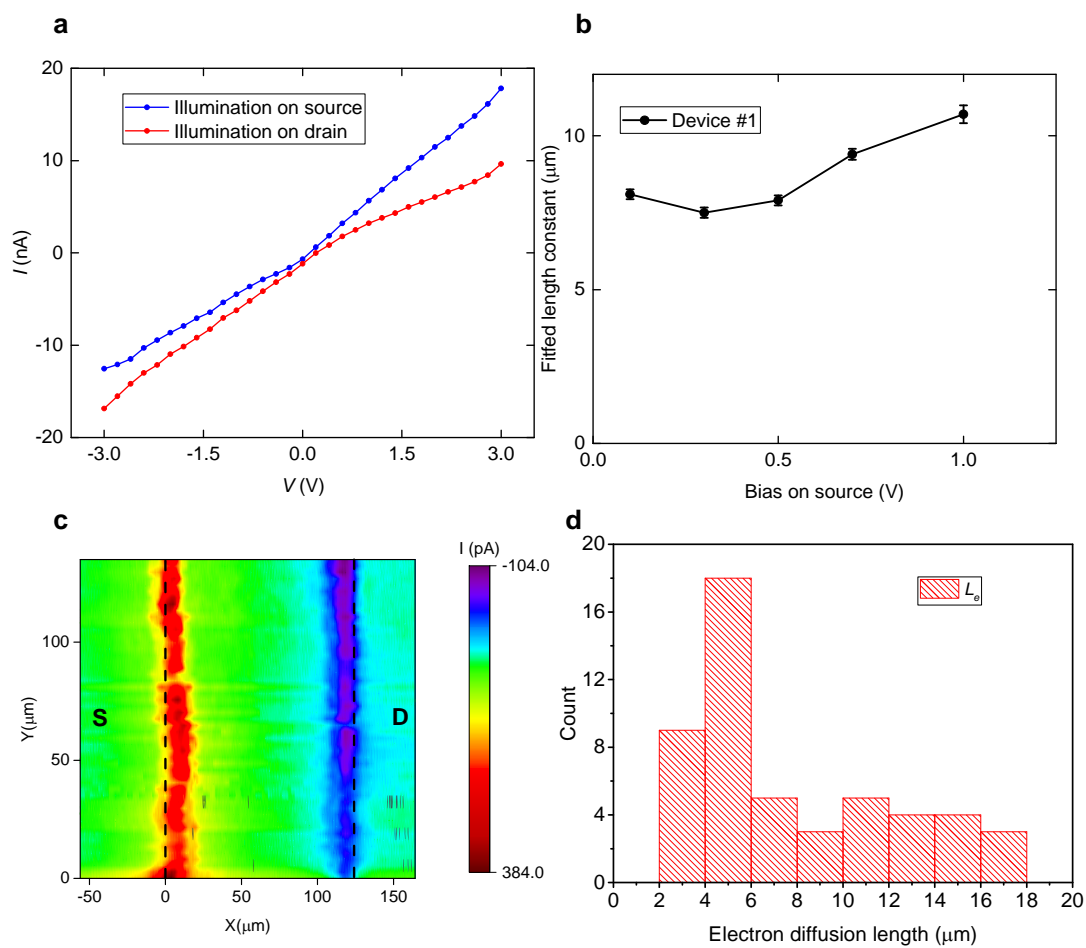

**Figure S4.** Measurement results for device #1 with 50 nm Au electrodes. (a) IV curves with only one contact being illuminated. Illumination intensity: 23.7 kW/m$^2$, illumination spot diameter: 25 μm and half of the spot is in the gap. (b) Fitted photocurrent decay length versus applied bias. (c) 2D scanning photocurrent image of device #1 under 0.1 V bias on source electrode. The two dashed lines indicate electrode edges. d) Histogram of electron diffusion length $L_e$ extracted from different line-scans in (a), yielding an average $L_e = 7.8 \pm 4.4$ μm. It can be seen from (c) that the top part of scanning has more rapid color change near source electrode edge on the left, which means this part of films has shorter diffusion length, possibly due to relatively poor film quality.

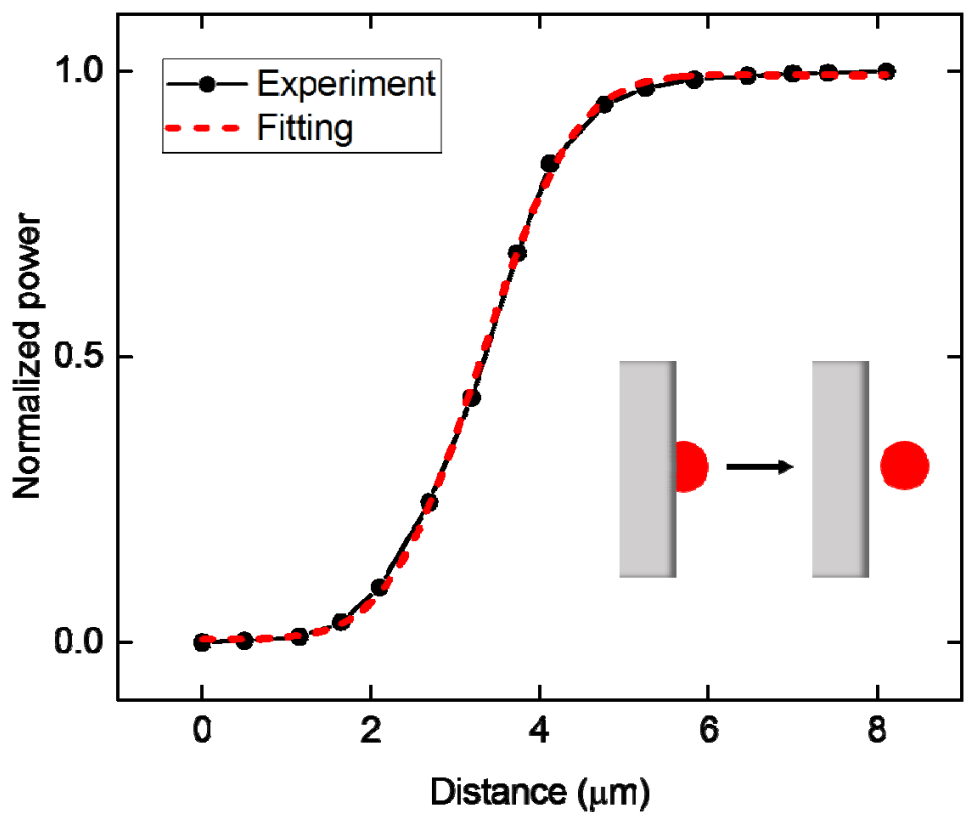

**Figure S5.** Laser spot diameter measurement. As the inset shows, the laser spot is focused onto the sharp edge of a razor blade. As the edge was scanned across the laser spot, the power of unblocked part of laser spot was recorded as a function of traveling distance. Fitted with an integral of Gaussian function, the width of the Gaussian beam is 1.7 μm.

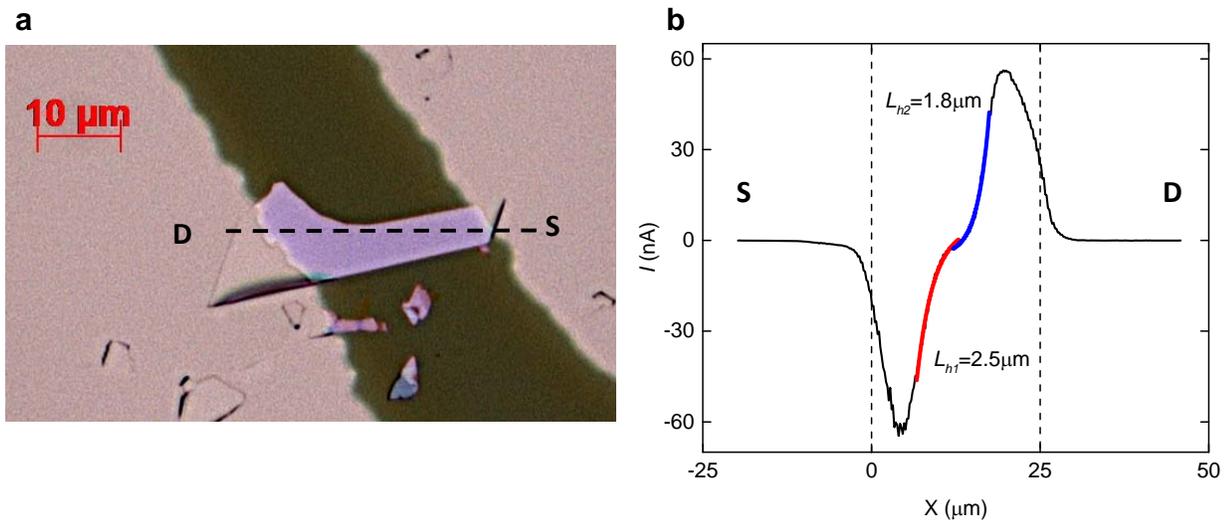

**Figure S6.** Scanning photocurrent measurement on a MoS$_2$ device. (a) Optical image of the multi-layer MoS$_2$ nanoflake device, the dashed line defines scanning track. (b) Typical line-scan with laser excitation, under 0 V source bias. Photocurrent decay length fitted from source side and drain side both show ~ 2 μm diffusion length, which is limited by the laser spot size.

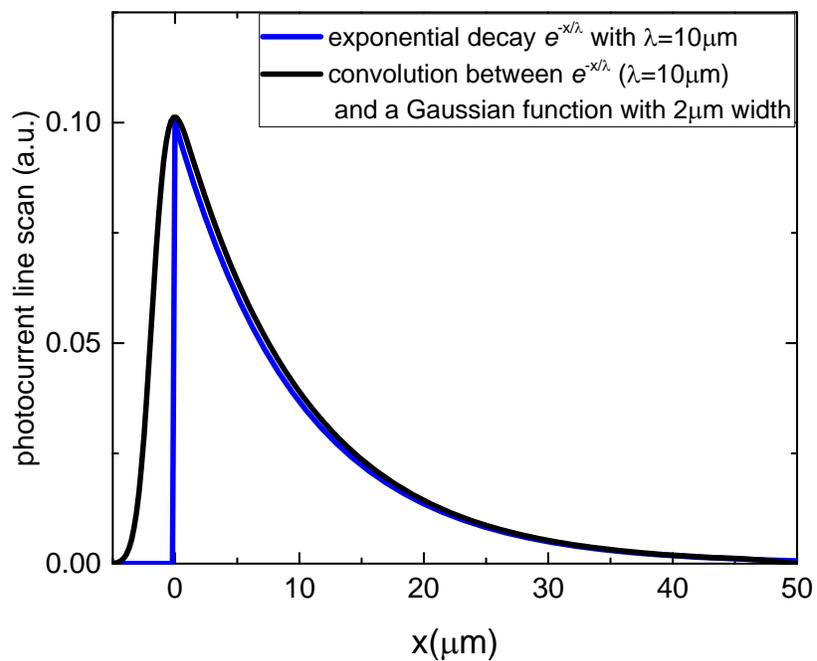

**Figure S7.** Comparison between a single exponential decay function $e^{-x/\lambda}$ and a convolution integral of Gaussian function with width 2 μm and $e^{-x/\lambda}$. $\lambda$ is set to 10 μm in $e^{-x/\lambda}$ in both functions. It is seen that although the convoluted curve shows rounded edge, the decreasing part at x>0 closely follows the behavior of simple exponential decay function $e^{-x/\lambda}$.

| Device # | Contact | Gap length (μm) | Experiments performed on the device | Fitted electron transport length (μm) |
|---|---|---|---|---|
| **#1** | 5nm Cr + 50nm Au | 124 | IV with different illumination intensities; IV with local illumination; 2D mapping; laser scanning with various biases | 7.8±4.4 at $V$= 0.1V (50 line-scans) |
| **#2** | 5nm Cr + 50nm Au | 138 | IV with local illumination | N/A |
| **#3** | 50nm Ni | 124 | IV with local illumination; big spot scanning | 7.3±1.8 at $V$=5V (2 line-scans) |
| **#4** | 80nm Ni | 120 | IV with different illumination intensities; IV with local illumination; 2D mapping; laser scanning with various biases | 10.5±1.6 $V$=0V (56 line-scans) 42.1±24.0 $V$=1V (7 line-scans) |
| **#5** | 5nm Cr + 50nm Au | 135 | IV with local illumination; big spot scanning | 5.0±1.6 at $V$=5V (4 line-scans) |
| **#6** | 80nm Ni | 126 | IV with different illumination intensities; IV with local illumination; laser scanning with various biases | 8.5±1.4 at $V$=0.1V (4 line-scans) 22.5±8.0 at $V$=1V (7 line-scans) |
| **#7** | 50nm Ni | 50 | IV with local illumination; big spot scanning | N/A |

**Table S1.** A list of detailed information (device number, contact type, channel length, experiments performed and carrier decay length extracted) for all the devices measured.